\documentclass{ar-1col-S2O}

\usepackage[comma]{natbib}
\usepackage{url}
\usepackage{color}
\usepackage{amsmath,amssymb,amsthm, bm}
\usepackage{mathtools}

\theoremstyle{definition}
\newtheorem{definition}{Definition}

\theoremstyle{plain}

\usepackage{mathtools}

\jname{Annu. Rev. Stat. Appl.}
\jvol{12}
\jyear{YYYY}
\doi{10.1146/((please add article doi))}

\begin{document}

\markboth{Kang et al.}{Invalid Instruments}

\title{Identification and Inference with Invalid Instruments}

\author{Hyunseung Kang,$^1$ Zijian Guo,$^2$ Zhonghua Liu,$^3$ and Dylan Small$^4$
\affil{$^1$Department of Statistics, University of Wisconsin-Madison, Madison WI, United States, 53706; email: hyunseung@stat.wisc.edu}
\affil{$^2$Department of Statistics, Rutgers University, Piscataway NJ, United States, 08854; email: zijguo@stat.rutgers.edu}
\affil{$^3$Department of Biostatistics, Columbia University, New York NY, United States, 10032; email: zl2509@cumc.columbia.edu}
\affil{$^4$Department of Statistics and Data Science, The Wharton School, University of Pennsylvania, Philadelphia PA, United States, 19104; email: dsmall@wharton.upenn.edu}}

\begin{abstract}
Instrumental variables (IVs) are widely used to study the causal effect of an exposure on an outcome in the presence of unmeasured confounding. IVs require an instrument, a variable that is (A1) associated with the exposure, (A2) has no direct effect on the outcome except through the exposure, and (A3) is not related to unmeasured confounders. Unfortunately, finding variables that satisfy conditions (A2) or (A3) can be challenging in practice. This paper reviews works where instruments may not satisfy conditions (A2) or (A3), which we refer to as invalid instruments. We review identification and inference under different violations of (A2) or (A3), specifically under linear models, non-linear models, and heteroskedatic models. We conclude with an empirical comparison of various methods by re-analyzing the effect of body mass index on systolic blood pressure from the UK Biobank.

\end{abstract}

\maketitle

\tableofcontents

\section{INTRODUCTION} \label{sec:intro}

\noindent Instrumental variables (IVs) are a popular tool to estimate the causal effect of an exposure on an outcome in the presence of unmeasured confounders, which are unmeasured variables that affect both the exposure and the outcome. Briefly, IVs require finding a variable called an instrument that satisfies three core conditions: 
\begin{enumerate}
\item[(A1)] The instrument is related to the exposure;
\item[(A2)] The instrument has no direct pathway to the outcome except through the exposure; 
\item[(A3)] The instrument is not related to unmeasured confounders that affect the exposure and the outcome. 
\end{enumerate} 
See \citet{hernan_instruments_2006,baiocchi_instrumental_2014} and Section \ref{sec:linmodel} for details. We remark that condition (A2) is often referred to as the exclusion restriction \citep{imbens1994identification,angrist_identification_1996}. IVs can be a valuable tool in settings where a randomized trial, which is the gold standard for dealing with unmeasured confounders, is impractical.

Oftentimes there is uncertainty about whether candidate instruments, in fact, satisfy (A1)-(A3), especially conditions (A2) and (A3); this problem is loosely referred to as the invalid instruments problem 
\citep{murray_avoiding_2006,conley_plausibly_2012}. For example, if the instrument is a genetic marker, which is the case in Mendelian randomization (MR) \citep{davey_smith_mendelian_2003,davey_smith_mendelian_2004}, satisfying (A2) would imply that the genetic marker's only biological function is to affect the exposure. However, this assumption is untenable for many genetic markers as they often have multiple biological functions, a phenomenon broadly known as pleiotropy \citep{solovieff2013pleiotropy,hemani2018evaluating}. More generally, without a complete understanding of the instruments' downstream effects on the outcome, IVs are plagued by possible violations of condition (A2). Also, when condition (A1) is weakly satisfied in that the instrument's association to the exposure is ``small''  (i.e. weak instruments; see \citet{staiger_instrumental_1997},\citet{stock_survey_2002} and \citet{andrews_weak_2019}), a slight violation of condition (A2) can lead to dramatically biased estimates of the causal effect of the exposure \citep{bound_problems_1995,hahn_estimation_2005,small_war_2008}.

Recent works have developed some promising frameworks to study the identification and inference of the causal effect of the exposure when some candidate instruments are, in fact, invalid. In this paper, we organize these works into the following categories:
\begin{enumerate}
\item \emph{Linear Models}: One of the earliest works in this literature is to generalize a popular linear, IV model to parametrize the violations of conditions (A2) and (A3). Roughly speaking, instruments are invalid under linear models if they have non-zero, partial effects on the outcome after conditioning on the exposure. Some works in this area include \citet{han_detecting_2008,kolesar2015identification,kang_instrumental_2016,guo_confidence_2018,windmeijer_lasso_2019,windmeijer_confidence_2021,fan2022endogenous}, and \cite{guo_causal_2023}. 
Given that the invalid instruments violate conditions (A2) and (A3) in a linear fashion, these works generate simple conditions for identification, most notably the majority rule. The majority rule states that a majority of the instruments are valid, without knowing which instruments are valid a priori. Also, some of these works have relatively straightforward methods for estimation, such as the median estimator. However, if the linear outcome model is mis-specified, these methods can lead to wrong conclusions about the effect of the exposure.



\item \emph{Nonlinear Models}: Recent works utilize the non-linearities in the exposure model when instruments do not satisfy conditions (A2) and (A3).For example, \cite{guo2022two} proposed finding non-linear trends in the exposure model via machine learning methods to identify and infer the causal effect of the exposure. 
\cite{sun2023semiparametric} proposed using higher-order interactions of instruments and G estimation \citep{robins1992estimating,robins1994correcting} to identify and infer the causal effect of the exposure. 
Critically, both of these methods rely on the exposure model's nonlinearity for identification, and it's important to check the required nonlinearity conditions to ensure that the identification conditions are plausible with the data \citep{lewbel2019identification}.

\item \emph{Heteroskedastic Models}: These works utilize heteroskedsaticity in the exposure model or the outcome model to identify the causal effect with invalid instruments. Some works in this area include \citet{lewbel2012using, tchetgen2017genius,sun2022selective,ye2021genius} and \citet{liu2023mendelian}. Similar to the nonlinear modeling framework above, these methods require heteroskedacity and the required heteroskedascity condition should be checked in practice.

\end{enumerate}
The rest of the paper goes into the details behind each category above. We also empirically compare the methods discussed in the paper by re-analyzing the causal effect of body mass index on blood pressure from the UK Biobank. We remark that the review does not discuss, in detail, an important interplay between weak instruments and invalid instruments and the challenges that they pose. The intersection between weak instruments and invalid instruments have recently gained considerable attention and we provide a summary of the recent works in this area plus other remaining questions in the study of invalid instruments at the end of the paper.

\section{LINEAR MODELS}

\subsection{Model and Definition of Invalid Instruments} \label{sec:linmodel}
Let $Y \in \mathbb{R}$ denote the outcome, $D \in \mathbb{R}$ denote the exposure, and $\mathbf{Z} \in \mathbb{R}^{p}$ denote the $p$ instruments. Consider the following model for $Y,D$ and $\mathbf{Z}$ where, without loss of generality, we assume that $Y$, $D$, and $\mathbf{Z}$ are centered to mean zero:
\begin{align} 
D &= \mathbf{Z}^\intercal \bm{\gamma} + \delta, && \mathbb{E}(\delta\mid \mathbf{Z} ) = 0, 
\label{eq:DLinModel} \\
Y &= D \beta + \mathbf{Z}^\intercal \bm{\pi} +  \epsilon, &&  \mathbb{E}(\epsilon \mid \mathbf{Z} ) = 0. 
\label{eq:YLinModel} 
\end{align}
In Equation \ref{eq:DLinModel}, 
the parameter $\bm\gamma \in \mathbb{R}^{p\times1}$ represents the instruments' relevance to the exposure. 
For this review, we only consider the case where all instruments are relevant (i.e., $\gamma_j \neq 0$ for all $j=1,\ldots,p$) in order to focus on invalid instruments.
For extensions to identifying and inferring the causal effect when some instruments are not relevant and some instruments are invalid, see \cite{guo_confidence_2018} and \citet{fan2022endogenous}.%

In Equation \ref{eq:YLinModel}, the parameter $\beta \in \mathbb{R}$ represents the effect of the exposure on the outcome and is the main parameter of interest in IVs. The parameter $\beta$ is sometimes referred to as a ``structural'' parameter \citep{goldberger1972structural,wooldridge_econometrics_2010,angrist2009mostly}, to distinguish it from the usual regression coefficient or a ``reduced-form'' parameter,  which can often be estimated consistently with ordinary least squares. Specifically, the estimate of the regression coefficients  from running an ordinary least squares regression of $Y$ on $D$ and $\mathbf{Z}$, denoted as $\hat{\beta}_{\rm ols}$, is generally inconsistent for $\beta$ because the variable $D$ is not independent of the error term $\epsilon $. In contrast, for the parameters in Equation \ref{eq:DLinModel}, we can run a ordinary least squares regression of $D$ on $\mathbf{Z}$ to obtain a consistent estimator of $\bm \gamma$.

The parameter $\bm \pi \in \mathbb{R}^{p \times 1}$ represents the effects of the instruments on the outcome after adjusting for the exposure. If $\bm \pi = \bm 0$, Equation \ref{eq:YLinModel} reduces to a well-studied, linear IV model  \citep{angrist2009mostly,wooldridge_econometrics_2010} and all $p$ instruments are said to be valid instruments. If $\bm \pi \neq \bm 0$, some of the $p$ instruments are not valid and the non-zero elements of $\bm \pi$ encode which of the instruments are invalid.
\begin{definition}[Valid Instrument] \label{def:validIV}
    Suppose Equation \ref{eq:DLinModel} and Equation \ref{eq:YLinModel} hold. Instrument $j \in \{1,\ldots,p\}$ is valid if $\pi_j=0$ and invalid if $\pi_j \neq 0$. Let $\mathcal{V}=\{j: \pi_j=0\}$ be the set of valid instruments. 
\end{definition}
If the set of valid instruments $\mathcal{V}$ is known a priori and there is at least one valid instrument (i.e., the size of the set, denoted as $|\mathcal{V}|$, is greater than or equal to $1$), Equation \ref{eq:YLinModel} again reduces to a well-studied linear IV model where the complement of $\mathcal{V}$ serves as the control variables; see Chapter 5 of \citet{wooldridge_econometrics_2010} for an example. But, in practice, the knowledge about $\mathcal{V}$ is unknown, and the central goal under the invalid IV framework is to study identification and inference of the causal effect when there is no a priori knowledge about which instruments among the $p$ candidate instruments are valid.

Definition \ref{def:validIV} of a valid instrument is closely related to the definition of a valid instrument under an additive, linear, constant effects (ALICE) potential outcomes model \citep{holland_causal_1988}. Let $Y(d,\mathbf{z}) \in \mathbb{R}$ be the potential outcome if an individual were to have exposure $d \in \mathbb{R}$ and instruments $\mathbf{z} \in \mathbb{R}^{p}$.  For $d, d' \in \mathbb{R}$ and $\mathbf{z}, \mathbf{z}' \in \mathbb{R}^{p}$, the ALICE model states
\begin{equation} \label{eq:ALICE}
Y(d',\mathbf{z}')  - Y(d,\mathbf{z}) = (d'-d) \tilde{\beta} + (\mathbf{z}' - \mathbf{z})^\intercal \tilde{\bm \psi}, \quad{} \mathbb{E}[Y(0,\mathbf{0})\mid \mathbf{Z}] = \mathbf{Z}^\intercal \tilde{\bm \phi}.
\end{equation}
The tilde notation in Equation \ref{eq:ALICE} emphasizes that the model parameters are from the potential outcomes model. The parameter $\tilde{\beta}$ represents the causal effect of changing the exposure by one unit on the outcome. Each $j$th element of $\tilde{\bm \psi} \in \mathbb{R}^{p \times 1}$ represents the causal effect of changing the $j$th instrument by one unit on the outcome. The term $\mathbb{E}[Y(0,\mathbf{0})\mid \mathbf{Z}] = \mathbf{Z}^\intercal \tilde{\bm \phi}$ where $\tilde{\bm \phi} \in \mathbb{R}^{p\times 1}$ represents the effect from unmeasured confounding. If the no direct effect condition (A2) is formalized as  $Y(d,\mathbf{z})=Y(d,\mathbf{z}')$ for all $d,\mathbf{z},\mathbf{z}'$, we have $\tilde{\bm\psi} = \mathbf{0}$. If the no instrument confounding condition (A3) is formalized as $\mathbf{Z} \perp Y(d,\mathbf{z})$ where $\perp$ stands for independence between two random variables, we have $\mathbf{Z}^\intercal \tilde{\bm \phi} = \mathbf{0}$. Also, if we assume the stable unit treatment value assumption \citep{rubin_comment_1980} or causal consistency where $Y = Y(D,\mathbf{Z})$, Equation \ref{eq:ALICE} simplifies to Equation \ref{eq:YLinModel} where $\tilde{\beta} = \beta$, $\bm\pi = \tilde{\bm\psi} + \tilde{\bm \phi}$, and $\epsilon = Y(0,\mathbf{0}) - \mathbb{E}[Y(0,\mathbf{0}) \mid \mathbf{Z}]$. In other words, under the ALICE model for potential outcomes and causal consistency, the violations of conditions (A2) and (A3) are summarized with the parameter $\bm \pi$. 

We make some other important remarks about the definition of a valid instrument. First, the validity of instrument $j$ depends on the candidate set of instruments. Instrument $j$ may be valid with one set of instruments but may not be valid with another set of instruments. 
Second, if we have covariates $\mathbf{X}$ that are independent of the error terms in Equation \ref{eq:DLinModel} and Equation \ref{eq:YLinModel}, we can adjust for $\mathbf{X}$ by first fitting a linear regression of $Y$ on $\mathbf{X}$, $D$ on $\mathbf{Z}$, and $\mathbf{Z}$ on $\mathbf{X}$. Then we replace $Y$, $D$, and $\mathbf{Z}$ in Equation \ref{eq:DLinModel} and Equation \ref{eq:YLinModel} with the residuals of the linear regressions from the first step. This procedure is justified by the Frisch-Waugh-Lovell theorem \citep{davidson_estimation_1993}. 
Third, several works \citep{hahn_estimation_2005,berkowitz2008nearly,berkowitz2012validity,guggenberger2012asymptotic,conley_plausibly_2012,armstrong2021sensitivity} have studied the properties of existing estimators and tests of $\beta$ when there is a near-violation of instrument validity. Roughly speaking, a near-violation of instrument validity is characterized as $\tilde{\bm \psi} = \mathbf{0}, \tilde{\bm \phi} = \mathbf{C}/\sqrt{n}$ or as $\bm \pi = \mathbf{C}/\sqrt{n}$ for some constant vector $\mathbf{C} \in \mathbb{R}^{p}$ and $n$ is the sample size. These works showed that existing estimators will be biased and tests of $\beta$ will have an inflated Type I error rate. 
Fourth, \cite{liao_adaptive_2013,cheng_select_2015,ditraglia2016using} and \citet{patel2024selection} considered estimating $\beta$ when there is a known set of valid instruments and another set of potentially invalid instruments. Finally, we remark that there are works on constructing bounds of $\beta$ \citep{small_sensitivity_2007,baiocchi_building_2010,kang_full_2016,ashley2015sensitivity,fogarty_biased_2021} 
under various assumptions about the magnitude of $\bm \pi$.

\subsection{Identification of the Causal Effect of the Exposure with Invalid Instruments} \label{sec:lin_ident}
To better understand how $\beta$ can be identified when the set of valid instruments $\mathcal{V}$ is unknown, it's useful to consider a model of $Y$ that only depends on $\mathbf{Z}$, often referred to as a reduced-form model: 
\begin{align}
\label{eq:Yreduced}
Y &= \mathbf{Z}^\intercal \bm\Gamma + e, && \bm \Gamma = \beta \bm \gamma + \bm \pi, \ e = \beta \delta + \epsilon, \ \mathbb{E}(e \mid \mathbf{Z}) = 0. 
\end{align}
With the observed data $Y,D$ and $\mathbf{Z}$, we can identify the parameters $\bm \gamma \in \mathbb{R}^{p \times 1}$ and $\bm \Gamma \in \mathbb{R}^{p \times 1}$ based on the following relationships from ordinary least squares:
\begin{equation*} 
\bm \Gamma = \mathbb{E}(\mathbf{Z}\mathbf{Z}^\intercal)^{-1} \mathbb{E}(\mathbf{Z} Y), \quad{} \bm \gamma = \mathbb{E}(\mathbf{Z} \mathbf{Z}^\intercal)^{-1} \mathbb{E}(\mathbf{Z} D). 
\end{equation*}
With the parameters $\bm \Gamma$ and $\bm \gamma$ identified from the data, we can reframe identifying $\beta$ as finding a unique value of $(\beta, \bm \pi)$ from $(\bm \gamma, \bm \Gamma)$ based on the system of linear equations in Equation \ref{eq:Yreduced} (i.e., $\bm \Gamma = \beta \bm \gamma + \bm \pi$).
Also, the system of linear equations reveals the role that $\mathcal{V}$ plays in identifying $\beta$. For example, suppose $\bm \pi = \mathbf{0}$ so that all instruments are valid and $\mathcal{V} = \{1,\ldots,p\}$. Then, the linear system of equations simplifies to $\bm \Gamma = \beta \bm \gamma$ and we can identify $\beta$ given $\bm \Gamma$ and $\bm \gamma$ by setting $\beta = \Gamma_j / \gamma_j$ for any $j = 1,\ldots, p$. Or, suppose the set of valid instruments $\mathcal{V}$ is known a priori and there is at least one valid instrument (i.e. $|\mathcal{V}| \geq 1$). Then, for $j \in \mathcal{V}$, the system of linear equations simplifies to $\Gamma_j = \beta \gamma_j$ and we can again identify $\beta$. Finally, if there are no restrictions on $\bm \pi$, there are no unique values of $(\beta, \bm \pi)$ given $(\bm \gamma, \bm \Gamma)$ since there are $p$ linear equations and $p+1$ unknown variables. 

When $\mathcal{V}$ is unknown, \citet{han_detecting_2008} and \citet{kang_instrumental_2016} proposed what's now called the majority rule to identify $\beta$:
\begin{equation}
|\mathcal{V}| > \frac{p}{2}.
\label{eq: majority}
\end{equation}
Simply put, the majority rule states that the number of valid instruments is more than 50\% of the instruments. Critically, we do not have to know a priori which instruments are valid; we simply have to know that the majority of the instruments are valid. 

We briefly illustrate how the majority rule in Equation \ref{eq: majority} places a constraint on the linear system of equations above to identify $\beta$; for the full proof, see \citet{kang_instrumental_2016} where they prove a necessary and sufficient condition to have a unique solution of $(\beta, \bm \pi)$ given $(\bm \gamma, \bm \Gamma)$ and a lower bound on $|\mathcal{V}|$. Given $(\bm \gamma, \bm \Gamma)$, let $(\beta, \bm \pi)$ and $(\beta', \bm \pi')$ be the solutions to the system of equations, i.e., $\bm \Gamma = \beta \bm \gamma + \bm \pi$ and $\bm \Gamma = \beta' \bm \gamma + \bm \pi'$. Let $\mathcal{V}$ and $\mathcal{V}'$ denote the sets of valid instruments defined by $\bm \pi$ and $\bm \pi'$, respectively. By the majority rule, $\mathcal{V} \cap \mathcal{V}' \neq \emptyset$ and we can always pick $j \in \mathcal{V} \cap \mathcal{V}'$ where $\pi_j= \pi_j' = 0$. For instrument $j$, the system of linear equations simplifies to $\Gamma_j = \beta \gamma_j$ and $\Gamma_j = \beta' \gamma_j$, which imply that $\beta = \beta'$. Furthermore, we have $\bm \pi = \bm \Gamma - \beta \bm \gamma = \bm \Gamma - \beta' \bm \gamma = \bm \pi'$ and thus, the solution to the system of linear equations is unique.

We can also construct a falsification test of the majority rule where the ``null hypothesis'' assumes that the majority rule holds. For example, suppose the majority rule holds, and the non-zero elements of $\bm \pi$ are far away from $0$. In this setting, there should only be one large cluster of instruments with the same $\Gamma_j / \gamma_j$, and the size of this cluster should be greater than $p/2$. If we do not see such a cluster, this indicates a violation of the majority rule. For additional details on how to conduct this test, see \citet{guo2023robust}.

Next, \citet{guo_confidence_2018} proposed 
the plurality rule to identify $\beta$: 
\begin{equation}
|\mathcal{V}|> \max_{c\neq 0} |\mathcal{I}_c| \quad \text{with} \quad \mathcal{I}_c=\left\{j \in \{1,\ldots,p\}: \frac{\pi_j}{\gamma_j} = c\right\}. 
\label{eq: plurality}
\end{equation}
In words, $\mathcal{I}_c$ denotes a subset of instruments that have a common value of $\Gamma_j/\gamma_j$, specifically $\Gamma_j /\gamma_j = \beta + c$. Note that the set of valid instruments $\mathcal{V}$ equals to $\mathcal{I}_0$ where $c = 0$. The plurality condition requires that the set of valid instruments $\mathcal{V}$ is the largest among all subsets of instruments with a common value of $\Gamma_j / \gamma_j$ that is not equal to $\beta$. Also, if the majority rule holds, the plurality rule holds since the size of the set $\mathcal{I}_c$ for any $c \neq 0$ cannot be greater than $p/2$, i.e., $p/2 > \max_{c \neq 0} |\mathcal{I}_c|$. In other words, the majority rule is a sufficient condition for the plurality rule. Finally, we can also falsify the plurality rule, albeit under more stringent conditions; see \citet{guo2023robust} for details. 

We conclude with a remark about the work by \citet{andrews_consistent_1999}. 
\citet{andrews_consistent_1999} primarily focused on developing a model selection procedure to consistently estimate $\mathcal{V}$. The model selection procedure relies on using a test statistic called the $J$ test \citep{hansen_large_1982} to distinguish between valid instruments and invalid instruments, and a couple of the works discussed below use this procedure to tune relevant tuning parameters. Also, when characterizing the properties of the proposed model selection procedure, \citet{andrews_consistent_1999} proposed a condition for identifying $\beta$ that is a version of the plurality rule. Specifically, for any subset $\mathcal{C} \subseteq \{1,\ldots,p\}$ and a vector $\mathbf{v} \in \mathbb{R}^p$, let $\mathbf{v}_{\mathcal{C}} \in \mathbb{R}^{|\mathcal{C}|}$ be the vector with elements defined by the subset $\mathcal{C}$. Then, \citet{andrews_consistent_1999} stated that $\beta$ is identified if 
\begin{equation*} 
\forall \mathcal{C} \text{ where } |\mathcal{C}| \geq |\mathcal{V}| 
\text{ and } \mathcal{V} \neq \mathcal{C}, \quad \bm\pi_{\mathcal{C}} \neq q \bm \gamma_{\mathcal{C}}
\end{equation*}
for any $q \neq 0$. 



\subsection{Estimation and Inference of the Causal Effect of the Exposure}
\subsubsection{Consistent Estimators of $\beta$} \label{sec:point_est}


We lay out the following notations to describe different estimators of $\beta$. For each study unit $i=1,\ldots,n$, let $(Y_i,D_i,\mathbf{Z}_i) \in \mathbb{R} \otimes \mathbb{R} \otimes \mathbb{R}^p$ be the observed outcome, exposure, and $p$ instruments, respectively. Let $\mathbf{Y} = (Y_1,\ldots,Y_n) \in \mathbb{R}^{n \times 1}$, $\mathbf{D}=(D_1,\ldots,D_n) \in \mathbb{R}^{n \times 1}$, and $\mathbf{Z} = (\mathbf{Z}_1,\ldots,\mathbf{Z}_n) \in \mathbb{R}^{n \times p}$. As before, without loss of generality, we assume that the vectors $\mathbf{Y}$ and $\mathbf{D}$ as well as the columns of the matrix $\mathbf{Z}$ are centered to mean zero. Let $\mathbf{P}_\mathbf{Z} = \mathbf{Z}(\mathbf{Z}^\intercal \mathbf{Z})^{-1} \mathbf{Z}^\intercal \in \mathbb{R}^{n \times n}$ be the projection matrix onto the column space of $\mathbf{Z}$ and let $\mathbf{P}_{\mathbf{Z}}^\perp = \mathbf{I} - \mathbf{\mathbf{P}}_{\mathbf{Z}} \in \mathbb{R}^{n \times n}$ be the residual projection matrix where $\mathbf{I} \in \mathbb{R}^{n \times n}$ is the identity matrix. For any vector $\mathbf{v} \in \mathbb{R}^p$ and $q\geq 1$, let $\|\mathbf{v}\|_q = (\sum_{j=1}^{p} v_j^q)^{1/q}$ be its $q$ norm and let $v_j$ denote the $j$th element of $\mathbf{v}$. Finally, for a set $\mathcal{V} \subseteq \{1,\ldots,p\}$, let $\mathcal{V}^c$ denote its complement and $\mathbf{Z}_{\mathcal{V}} \in \mathbb{R}^{n \times |\mathcal{V}|}$ denote the matrix $\mathbf{Z}$ with the columns specified by $\mathcal{V}$.

We start by describing the two-stage least squares (TSLS) estimator, a popular estimator of $\beta$ when $\mathcal{V}$ is known a priori and $|\mathcal{V}| \geq 1$. The two-stage least squares estimator first fits a linear regression model of $\mathbf{D}$ on $\mathbf{Z}$ and obtains the predicted values of $\mathbf{D}$, i.e., $\mathbf{P}_\mathbf{Z} \mathbf{D}$. Second, it fits a linear regression model of $\mathbf{Y}$ on $\mathbf{P}_\mathbf{Z} \mathbf{D}$ and $\mathbf{Z}_{\mathcal{V}^c}$. If $\mathcal{V}^c$ is an empty set (i.e., all of the instruments are valid), we drop the $\mathbf{Z}_{\mathcal{V}^c}$ term in the second regression model. The two-stage least squares estimator of $\beta$ is the estimated regression coefficient in front of $\mathbf{P}_\mathbf{Z} \mathbf{D}$ from the second regression model. More succinctly, given a set of valid IVs $\mathcal{V}$ and $|\mathcal{V}|\geq 1$, the two-stage least squares is the solution to the following optimization problem: 
\begin{equation} \label{eq:tsls_est}
\bigl[\hat{\beta}_{\rm tsls}(\mathcal{V}), \hat{\bm \pi}_{\rm tsls}(\mathcal{V})\bigr]= {\rm argmin}_{\beta, \bm \pi_{\mathcal{V}^c}} \frac{1}{2} \| \mathbf{P}_\mathbf{Z}(\mathbf{Y} - \mathbf{D} \beta - \mathbf{Z}_{\mathcal{V}^c} \bm \pi_{\mathcal{V}^c}) \|_2^2. 
\end{equation}
Some works on invalid instruments compare their proposed estimators of $\beta$, which do not know $\mathcal{V}$ a priori, with the two-stage least squares estimator with a known $\mathcal{V}$; if used in this context, the two-stage least squares estimator is sometimes referred to as the oracle estimator \citep{guo_confidence_2018,windmeijer_confidence_2021}. If their proposed estimator of $\beta$ is asymptotically equivalent to the oracle estimator, the proposed estimator is said to be oracle-optimal in the literature. 

One of the first methods to estimate $\beta$ when $\mathcal{V}$ is unknown a priori is the median estimator of \citet{han_detecting_2008}. Consider the ordinary least squares estimators of $\bm \Gamma$ and $\bm \gamma$:
\begin{equation*} 
\hat{\bm \Gamma} = (\mathbf{Z}^\intercal \mathbf{Z})^{-1} \mathbf{Z}^\intercal \mathbf{Y}, \quad{} \hat{\bm \gamma} = (\mathbf{Z}^\intercal \mathbf{Z})^{-1} \mathbf{Z}^\intercal \mathbf{D}.
\end{equation*}
Under mild assumptions (see Chapter 4 of \citet{wooldridge_econometrics_2010}), $\hat{\bm \Gamma}$ and $\hat{\bm \gamma}$ are asymptotically normal:
 \begin{equation} \label{reduced form conv}
 \sqrt{n}\left[\begin{pmatrix}
   \hat{\bm \Gamma} \\ \hat{\bm \gamma}
 \end{pmatrix} - \begin{pmatrix}
   \bm \Gamma \\ \bm \gamma
 \end{pmatrix}\right] \xrightarrow{d} N\left[\begin{pmatrix}
   \bm 0 \\ \bm 0
 \end{pmatrix}, \begin{pmatrix} \bm \Omega_{\Gamma}
   & \bm \Omega_{\Gamma \gamma} \\ \bm \Omega_{\Gamma \gamma}^\intercal & \bm\Omega_{\gamma}
   \end{pmatrix}\right].
\end{equation}
We remark that the covariance matrices $\bm \Omega_{\Gamma} \in \mathbb{R}^{p \times p}, \bm \Omega_{\Gamma \gamma} \in \mathbb{R}^{p \times p}$ and $\bm \Omega_{\gamma} \in \mathbb{R}^{p \times p}$ can be consistently estimated. The median estimator of $\beta$, denoted as $\hat{\beta}_{\rm med}$, can be written as the median of $p$ ratios of $\hat{\Gamma}_j/\hat{\gamma}_j$, $j=1,\ldots,p$:
\begin{align} \label{eq:med_est}
\hat{\beta}_{\rm med} &= {\rm median} \{ \tilde{\beta}_1,\ldots,\tilde{\beta}_{p} \}, \quad{} \tilde{\beta}_j = \frac{\hat{\Gamma}_j}{\hat{\gamma}_j}.
\end{align}
\citet{han_detecting_2008} established that $\hat{\beta}_{\rm med}$ is consistent if the majority rule in Equation \ref{eq: majority} holds. Later, \citet{windmeijer_lasso_2019} established that the limiting distribution of the median estimator is an order statistic of a normal distribution. However, inference based on the median estimator is challenging due to the non-negligible bias of the order statistic \citep{windmeijer_lasso_2019, guo2023robust}.

\citet{kang_instrumental_2016} proposed a Lasso-based estimator of $\beta$, which the authors referred to as SISVIVE (Some Invalid, Some Valid Instrumental Variables Estimator). SISVIVE is inspired by the two-stage least squares estimator in Equation \ref{eq:tsls_est} and directly solves for the model parameters in Equation \ref{eq:YLinModel} with a penalty term on $\bm \pi$: 
\begin{equation} 
\label{eq:sisVIVE}
\bigl[ \hat{\beta}_{\rm sisvive}, \hat{\bm \pi}_{\rm sisvive} \bigr] =  \underset{\beta, \bm \pi}{\rm argmin} \ \frac{1}{2} \|\mathbf{P}_\mathbf{Z} (\mathbf{Y} - \mathbf{D}\beta - \mathbf{Z} \bm\pi)\|_2^2 + \lambda \|\bm \pi\|_1, \quad{} \lambda > 0.
\end{equation}
 This optimization problem can be solved with existing penalized regression software by reformulating Equation \ref{eq:sisVIVE} as follows:
\begin{align*}
\hat{\bm \pi}_{\rm sisvive} &=  \underset{ \bm \pi}{\rm argmin} \ \frac{1}{2} \| (\mathbf{P}_{\mathbf{Z}}- \mathbf{P}_{\mathbf{P}_\mathbf{Z} \mathbf{D}})(\mathbf{Y} - \mathbf{Z}\bm \pi)\|_2^2 + \lambda \| \bm \pi\|_1, \\
\hat{\beta}_{\rm sisvive} &= \frac{(\mathbf{P}_\mathbf{Z} \mathbf{D})^\intercal (\mathbf{Y}-\mathbf{Z}\hat{\bm \pi}_{\rm sisvive})}{ \|\mathbf{P}_{\mathbf{Z}} \mathbf{D}\|_2^2}.  
\end{align*}
The first step of the two-step procedure estimates $\bm \pi$ by using existing software for the Lasso (e.g. \citet{efron2004least}). The second step is a dot product between two $n$ dimensional vectors.  \citet{kang_instrumental_2016} established conditions when $\hat{\beta}_{\rm sisvive}$ is consistent for $\beta$ and recommended choosing the tuning parameter $\lambda$ by  cross-validation. Later, \citet{Bao_StatMed_2019} studied the finite sample properties of $\hat{\beta}_{\rm sisvive}$ through a simulation study.

Finally, \citet{kolesar2015identification} showed that the $k$-class estimator from \citet{anatolyev2013instrumental}, denoted as $\hat{\beta}_{\rm kclass}$ and formalized as
\begin{equation*}
\hat{\beta}_{\rm kclass} = \frac{\mathbf{D}^\intercal (\mathbf{I} - k \mathbf{P}_{\mathbf{Z}}^\perp ) \mathbf{Y}}{\mathbf{D}^\intercal (\mathbf{I} - k \mathbf{P}_{\mathbf{Z}}^\perp) \mathbf{D}}, \quad{} k = \frac{1 - \frac{1}{n}}{1 - \frac{p}{n} - \frac{1}{n}}, 
\end{equation*}
is consistent for $\beta$ when both the number of instruments $p$ and the number of samples $n$ grow to infinity and 
the following orthogonality condition between $\bm \pi$ and $\bm \gamma$ hold: 
\begin{equation} \label{eq:ortho}
\frac{1}{n} \bm \pi^\intercal \mathbf{Z}^\intercal \mathbf{Z} \bm \gamma \to 0, \quad{} \frac{p}{n} \to c \in [0,1).
\end{equation}
In words, the orthogonality condition states that the effect of the instruments on the exposure (i.e., $\bm \gamma$) is orthogonal to the direct effect of the instruments on the outcome (i.e., $\bm \pi$). If all the instruments are mutually independent of each other, the orthogonality condition roughly translates to $\bm \pi^\intercal \bm \gamma \approx 0$. Under this case, the system of linear equations in Equation \ref{eq:Yreduced} can be rewritten as $\bm \Gamma^\intercal \bm \gamma \approx \beta \bm \gamma^\intercal \bm \gamma$ and the parameter $\beta$ can be identified given $(\bm \gamma, \bm \Gamma)$. 

Unfortunately, beyond consistency, $\hat{\beta}_{\rm med}$ and $\hat{\beta}_{\rm sisvive}$ have no inferential guarantees, such as having a limiting normal distribution to enable testing $H_0: \beta = 0$ or constructing a confidence interval for $\beta$. Also, to establish asymptotic normality of $\hat{\beta}_{\rm kclass}$, \citet{kolesar2015identification} further assumed that the parameters $\bm \gamma$ and $\bm \pi$ are random and follow a multivariate normal distribution. The next two sections discuss some progress on conducting inference about $\beta$.

\subsubsection{Pointwise Inference of $\beta$}
\label{sec: pointwise inference}

We start with \citet{windmeijer_lasso_2019}, who proposed to use an adaptive version of the SISVIVE estimator. Specifically, consider the adaptive Lasso \citep{zou2006adaptive} version of the SISVIVE estimator in Equation \ref{eq:sisVIVE} where the initial estimator is the median estimator in Equation \ref{eq:med_est}:
\begin{align*}
\hat{\bm \pi}_{\rm med} &= \hat{\bm \Gamma} - \hat{\bm \gamma} \hat{\beta}_{\rm med}, \\
\hat{\bm \pi}_{\rm adlasso} &= \underset{ \bm \pi}{\rm argmin } \ \frac{1}{2} \| (\mathbf{P}_{\mathbf{Z}} - \mathbf{P}_{\mathbf{P}_\mathbf{Z} \mathbf{D}})(\mathbf{Y} - \mathbf{Z} \bm\pi)\|_2^2 + \lambda  \sum_{j=1}^{p} \frac{1}{|\hat{\pi}_{{\rm med},j}|} |\pi_j|, \quad{} \lambda > 0\\
\hat{\beta}_{\rm adlasso} &= \frac{(\mathbf{P}_{\mathbf{Z}} \mathbf{D})^\intercal (\mathbf{Y}-\mathbf{Z}\hat{\bm \pi}_{\rm adp})}{ \|\mathbf{P}_{\mathbf{Z}} \mathbf{D}\|_2^2}.  \label{eq:sisVIVE_2}
\end{align*}
\citet{windmeijer_lasso_2019} showed that  $\hat{\beta}_{adlasso}$ is consistent and asymptotically normal if the majority rule holds. \citet{windmeijer_lasso_2019} also proposed a method to choose $\lambda$ by using a downward testing procedure of \citet{andrews_consistent_1999}.

\citet{guo_confidence_2018} proposed a different approach, called two-stage hard thresholding (TSHT), to conduct inference on $\beta$. Broadly speaking, TSHT treats each instrument $j$ as a voter and uses a plurality voting procedure to estimate $\beta$. Specifically, consider a voting matrix $\mathbf{H} \in \mathbb{R}^{p \times p}$ where $H_{j,k}=1$ if the pair of instruments  $j$ and $k$ yield similar estimates of $\beta$ and $H_{j,k} =0$ if the pair yield different estimates of $\beta$, i.e.,
\begin{equation}
H_{j,k} = \begin{cases} 1 & \text{ if } \left| \frac{\hat{\Gamma}_j}{\hat{\gamma}_j}-\frac{\hat{\Gamma}_k}{\hat{\gamma}_k}\right| \leq z_{1-\frac{\alpha}{2p}} \cdot \hat{\rm se}\left(\frac{\hat{\Gamma}_j}{\hat{\gamma}_j}-\frac{\hat{\Gamma}_k}{\hat{\gamma}_k} \right), \\
0& \text{otherwise}.
\end{cases}
\label{eq: voting matrix}
\end{equation}
The term $\alpha\in (0,1)$ is the pre-specified significance level, $z_{1-\alpha/2p}$ is the $1-\alpha/2p$ quantile of the standard normal distribution, and $\hat{\rm se}(\hat{\Gamma}_j/\hat{\gamma}_j-\hat{\Gamma}_k/\hat{\gamma}_k)$
is the estimated standard error of the difference between $\hat{\Gamma}_j/\hat{\gamma}_j$ and $\hat{\Gamma}_k/\hat{\gamma}_k$. 
For each $j =1,\ldots,p$, let $\|\mathbf{H}_{j,\cdot}\|_0 = \sum_{k=1}^{p} H_{j,k}$ denote the number of non-zero elements in the $j$th row of $\mathbf{H}$ (i.e. $\mathbf{H}_{j,\cdot}$). Roughly speaking, $\|\mathbf{H}_{j,\cdot}\|_0$ 
measures the number of instruments that are close to the $j$th instrument's ratio $\hat{\Gamma}_j /\hat{\gamma}_j$. $\|\mathbf{H}_{j,\cdot}\|_0$ can also be thought of as the number of votes that instrument $j$ received on being the valid instrument. From $\|\mathbf{H}_{j,\cdot}\|_0$, we can estimate the set of valid instruments by picking instruments that received a majority or a plurality of votes: 
\begin{equation*}
    \hat{\mathcal{V}}_{\rm tsht}:=\left\{j \in \{1,\ldots,p\} \ \bigl| \ \|\mathbf{H}_{j,\cdot}\|_0 > \frac{p}{2} \right\} \cup 
    \left\{j \in \{1,\ldots,p\} \ \bigl| \  \|\mathbf{H}_{j,\cdot}\|_0 = \max_{k} \|\mathbf{H}_{k,\cdot}\|_0 \right\} 
    \label{vhat mp}
\end{equation*}  
After estimating $\mathcal{V}$, we can use the two-stage least squares estimator with the set $\hat{\mathcal{V}}_{\rm tsht}$ to estimate $\beta$. Under the plurality rule, \citet{guo_confidence_2018} showed that this estimator, denoted as $\hat{\beta}_{\rm tsht}$, is consistent, asymptotically normal and oracle-optimal for $\beta$.  \cite{zhang_fighting_2022} recently proposed an improvement of TSHT that prevents choosing a large number of irrelevant instruments through a resampling method. They showed that their procedure is effective in regimes where the number of instruments $p$ is very large.

\citet{windmeijer_confidence_2021} proposed the confidence interval method (CIM) for conducting inference on $\beta$. Roughly speaking, CIM uses ``working'' confidence intervals of $\beta$ to cluster instruments and picks the largest cluster of instruments with overlapping confidence intervals. Specifically, for each instrument $j$, CIM first constructs $p$ working confidence interval ${\rm CI}_j(q_n)$ of $\beta$:
\begin{align*}
 {\rm CI}_j(q_n) = [\hat{\beta}_j - q_n \cdot \hat{\rm se}(\hat{\beta}_j), \hat{\beta}_j +q_n \cdot  \hat{\rm se}(\hat{\beta}_j)], \quad{} \tilde{\beta}_j &= \frac{\hat{\Gamma}_j}{\hat{\gamma}_j}. 
\end{align*} 
The term $\hat{\rm se}(\hat{\beta}_j)$ is the standard error of $\tilde{\beta}_j$ based on applying the delta method to Equation \ref{reduced form conv}. The parameter $q_n$, which depends on the sample size $n$, is set to measure the similarity between confidence intervals. Note that $q_n$ is not equal to  $z_{1-\alpha/2}$, the $1-\alpha/2$ quantile of the standard normal distribution. Instead, \citet{windmeijer_confidence_2021} proposed an adaptive approach to set $q_n$ based on a downward testing procedure of \citet{andrews_consistent_1999}.  Second, the procedure constructs $K \leq p$ subgroups of instruments where all instruments in the subgroup have overlapping working confidence intervals:
\[
V_{k} = \left\{j  \mid {\rm CI}_j(q) \cap {\rm CI}_{j'} (q) \neq \emptyset \ \forall j,j' \in \{1,\ldots,p\} \right\} ,\quad{} k=1,\ldots,K.
\]
The CIM estimator of $\mathcal{V}$ is the largest subset of instruments where all the confidence intervals in the subset overlap with each other: 
\[
\hat{\mathcal{V}}_{\rm cim}= \{ V_{k} \mid |V_{k}| = \max_{k'=1,\ldots,K} |V_{k'}| \}.
\]
\citet{windmeijer_confidence_2021} show that if we construct a two-stage least squares estimator of $\beta$ with $\hat{\mathcal{V}}_{\rm cim}$, the estimator, denoted as $\hat{\beta}_{\rm cim}$, is consistent, asymptotically normal, and oracle-optimal. 

While both TSHT and CIM produce confidence intervals for $\beta$ and are oracle-optimal, a major downside of both procedures is that they rely on correctly estimating the set of valid instruments asymptotically, i.e., $\lim_{n \to \infty} P(\hat{\mathcal{V}}=\mathcal{V}) = 1$; this property is sometimes referred to as selection consistency. As noted in the post-selection inference literature (e.g. \citet{leeb2005model} and \citet{berk2013valid}), 
relying on selection consistency to enable inference on $\beta$ can lead to poor finite-sample properties, such as inflated Type I errors. This phenomenon is exacerbated when the true $\bm \pi$ is close to $\mathbf{0}$ so that selection consistency is not guaranteed. The next section highlights some progress in this area by constructing uniformly valid confidence intervals of $\beta$.

\subsubsection{Uniformly Valid Inference of $\beta$} 
\label{sec: uniform CI}


We discuss two procedures that construct uniformly valid confidence intervals of $\beta$.
\citet{kang2022two} proposed to take a union of several confidence intervals of $\beta$ constructed from subsets of instruments that pass the $J$ test \citep{hansen_large_1982}. Specifically,  suppose the investigator believes that at least $v \geq 1$ instruments are valid (i.e., $|\mathcal{V}| \geq v$) and wants to construct a $1-\alpha$ confidence interval of $\beta$. The union confidence interval, denoted as ${\rm CI}_{\rm union}$, takes a union of $1-\alpha_t$ confidence intervals of $\beta$ that use $v$ instruments and the $J$ test does not reject the null with the $v$ instruments at level $\alpha_s$: 
\[
{\rm CI}_{\rm union}^{[v]} = \cup_{\mathcal{V}', |\mathcal{V}'| =v} \left\{ {\rm CI}_{1 -\alpha_t}(\mathcal{V}') \mid J(\mathcal{V}') \leq q_{1-\alpha_s} \right\}, \quad{} \alpha = \alpha_s + \alpha_t, \quad{} |\mathcal{V}| \geq v \geq 1.
\]
Here, ${\rm CI}_{1-\alpha_t}(\mathcal{V}')$ is the $1-\alpha_t$ confidence interval of $\beta$ using $\mathcal{V}'$ as valid instruments, $J(\mathcal{V}')$ is the $J$ test using $\mathcal{V}'$ as valid instruments, and $q_{1-\alpha_s}$ is the $1-\alpha_s$ quantile of the $J$ test under its null hypothesis. The confidence interval ${\rm CI}_{1-\alpha_t}$ can be any confidence interval of $\beta$ that has the desired $1-\alpha_t$ coverage if valid instruments are used. For example, ${\rm CI}_{1-\alpha_t}$ can be the Wald confidence interval from the two-stage least squares estimator in Equation \ref{eq:tsls_est}, the Anderson-Rubin confidence interval \citep{anderson_estimation_1949}, or the confidence interval based on inverting the conditional likelihood ratio test \citep{moreira_conditional_2003}. Also, the terms $\alpha_s$ and $\alpha_t$ satisfy the constraint $\alpha_s + \alpha_t = \alpha$. For instance, choosing $\alpha_s = 0.01$ and $\alpha_t = 0.04$ would lead to a $95$\% confidence interval for ${\rm CI}_{\rm union}^{[v]} $. A main advantage of the union confidence interval is that it does not rely on selection consistency and is guaranteed to yield a $1-\alpha$ confidence interval of $\beta$.  However, the procedure requires an exponential number of computations and is generally infeasible for a moderate to a large number of instruments.

 \citet{guo_causal_2023} proposed another approach to construct a uniformly valid confidence interval of $\beta$. To illustrate the main idea, we focus on the case where the majority rule holds; for the case under the plurality rule, see \citet{guo_causal_2023}. \citet{guo_causal_2023} proposed the searching confidence interval of $\beta$ based on the relationship between $\Gamma$ and $\gamma$ in Equation \ref{eq:Yreduced}:
\begin{equation} \label{eq: thresholding searching}
     {\rm CI}_{\rm search}= \left\{\beta \in \mathbb{R}: L(\beta)>\frac{p}{2} \right\}, \quad{}
     L(\beta)=\sum_{j = 1}^p I\left[|\hat{\Gamma}_j-\beta\hat{\gamma}_j|\leq 
z_{1-\frac{\alpha}{2p}} \cdot \hat{\rm se} (\hat{\Gamma}_j-\beta\hat{\gamma}_j) \right].
\end{equation}
The term $\hat{\rm se} (\hat{\Gamma}_j-\beta\hat{\gamma}_j)$ is the estimated standard error of $\hat{\Gamma}_j-\beta\hat{\gamma}_j$ using the delta method, and $I[\cdot]$ is the indicator function. The term $L(\beta)$ measures the number of valid instruments for a particular value of $\beta$, and the interval ${\rm CI}_{\rm search}$ collects all $\beta$ values that lead to more than $50\%$ of instruments being valid.  
Also, to reduce the length of ${\rm CI}_{\rm search}$, \citet{guo_causal_2023} proposed a sampling version of ${\rm CI}_{\rm search}$, denoted as ${\rm CI}_{\rm ss}$. The sampling confidence interval starts by re-sampling $(\hat{\bm \Gamma}, \hat{\bm \gamma})$ $M$ number of times based on Equation \ref{reduced form conv}:
 \begin{equation*} 
     \begin{pmatrix}
     \hat{\bm \Gamma}^{[m]} \\
     \hat{\bm \gamma}^{[m]}
     \end{pmatrix} \overset{iid}{\sim}
     N\left[
     \begin{pmatrix}
     \hat{\bm \Gamma} \\
     \hat{\bm \gamma}
     \end{pmatrix},
     \frac{1}{n}\begin{pmatrix}
     \hat{\bm \Omega}_{\Gamma} & \hat{\bm \Omega}_{\Gamma \gamma}  \\
     \hat{\bm \Omega}_{\Gamma \gamma}^\intercal  & \hat{\bm \Omega}_{\gamma}      \end{pmatrix}
     \right], \quad \quad 1\leq m\leq M,
 \end{equation*}
where the terms $\hat{\bm \Omega}_{\Gamma}, \hat{\bm \Omega}_{\Gamma\gamma}$, and $\hat{\bm \Omega}_{\gamma}$ denote consistent estimators of 
${\bm \Omega}_{\Gamma}, {\bm \Omega}_{\Gamma\gamma}$, and ${\bm \Omega}_{\gamma}$, respectively.
 We then recompute the searching confidence interval in Equation \ref{eq: thresholding searching} from the resampled $(\hat{\bm \Gamma}^{[m]}, \hat{\bm \gamma}^{[m]})$:
{\small \begin{equation}
{\rm CI}^{[m]}=\left\{\beta\in \mathbb{R}: L^{[m]}(\beta)> \frac{p}{2}\right\}, \; {L}^{[m]}(\beta)=\sum_{j=1}^p I\left[|\hat{\Gamma}_j^{[m]}-\beta\hat{\gamma}_j^{[m]}|\leq \lambda \cdot z_{1-\frac{\alpha}{2p}} \cdot \hat{\rm se} (\hat{\Gamma}_j-\beta\hat{\gamma}_j)\right].
 \label{eq: shrink thre}
 \end{equation}}
The main difference between Equation \ref{eq: thresholding searching} and Equation \ref{eq: shrink thre} is the shrinkage parameter $\lambda$ in Equation \ref{eq: shrink thre} with  $\lambda=c_n\left( M^{-1}\log n\right)^{\frac{1}{2p}}$ where $c_n$ is a data-dependent parameter; see \citet{guo_causal_2023} for details. 
The sampling confidence interval aggregates non-empty intervals of ${\rm CI}^{[m]}$ by taking the lower and the upper limit of the interval ${\rm CI}^{[m]}$, denoted as $l^{[m]}$ and $u^{[m]}$, respectively:
 \begin{equation*} 
{\rm CI}_{\rm ss}=\left(\min_{m\in \mathcal{M}}l^{[m]},\max_{m\in \mathcal{M}}u^{[m]}\right),  \quad \mathcal{M}=\left\{1 \leq m \leq M : {\rm CI}_{\rm search}^{[m]}\neq\emptyset \right\}.
\end{equation*}
\citet{guo_causal_2023} showed that both ${\rm CI}_{\rm search}$ and ${\rm CI}_{\rm ss}$ achieve the desired coverage level even in the presence of instrumental variable selection error. Also, the length of both intervals are of the order $1/\sqrt{n}.$ In finite samples, we observe that ${\rm CI}_{\rm search}$ and ${\rm CI}_{\rm ss}$ are longer than the confidence intervals from TSHT and CIM, which is to be expected since ${\rm CI}_{\rm search}$ and ${\rm CI}_{\rm ss}$ guarantee uniform coverage of $\beta$.


\subsection{Connection to Two-Sample Summary Data Design in Mendelian Randomization}
 Linear models often serve as the data generating model for a popular study design in Mendelian randomization (MR) called the two-sample, summary data design \citep{pierce_efficient_2013, burgess_mendelian_2013}. In this section, we briefly discuss the connection between the assumptions underlying two-sample, summary data designs in Mendelian randomization and the assumptions discussed in the prior sections. For a full review of assumptions underlying MR, see \citet{bowden_framework_2017,slob2020comparison} and \citet{sanderson2022mendelian}.

Briefly, two-sample, summary data designs assume that the data is generated from two independent samples and only summary statistics, usually estimates of $\hat{\bm \Gamma}$ and $\hat{\bm \gamma}$ along with their corresponding standard errors, are available. The summary statistics are assumed to follow a multivariate normal distribution with diagonal covariance matrices. These assumptions are formalized below: 
 \begin{equation}   \label{eq:summaryMR}
 \bm \Gamma = \beta \bm\gamma + \bm \pi, \quad{}\quad{} \hat{\Gamma}_j \overset{\text{ind}}{\sim} N(\Gamma_j, \sigma_j^2), \ \hat{\gamma}_j \overset{\text{ind}}{\sim} N(\gamma_j, \omega_j^2 ), \  \hat{\bm \Gamma} \perp \hat{\bm \gamma}.
\end{equation}
See \citet{bowden_framework_2017, zhao_statistical_2020} and \citet{ye2021debiased} for additional details. In terms of identification, two-sample summary data designs assume the same relationship $\bm \Gamma = \beta \bm \gamma  + \bm \pi$ in Equation \ref{eq:Yreduced}. Also, many works in two-sample summary data Mendelian randomization make similar assumptions about $\bm \pi$ as those in Section \ref{sec:lin_ident}. For example, \citet{bowden2016consistent} proposed the weighted median estimator of $\beta$, which is consistent whenever the majority rule in Equation \ref{eq: majority} holds. \citet{hartwig2017robust} proposed the Zero Modal Pleiotropy Assumption (ZEMPA), which is the MR version of the plurality rule in Equation \ref{eq: plurality}, and showed that their proposed modal estimator of $\beta$ is consistent. An improvement of the modal estimator was proposed by \citet{burgess2018modal}. \citet{yao2023robust} proposed MR-SPI as a modified version of the methods proposed in \citet{guo_confidence_2018} and \citet{guo_causal_2023}. \citet{bowden2015mendelian} proposed the Instrument Strength Independent of Direct Effect (InSIDE) assumption, which is the MR version of the orthogonality condition in Equation \ref{eq:ortho}. We remark that the InSIDE assumption is similar to balanced horizontal pleiotropy \citep{bowden_framework_2017,hemani2018evaluating,zhao_statistical_2020}, which states that $\pi_j \overset{\text{ind}}{\sim}  N(0, \tau^2)$ and $\pi_j$'s are independent of $\hat{\bm \Gamma}$ and $\hat{\bm \gamma}$.

In terms of inference, two-sample summary data designs in Equation \ref{eq:summaryMR} place stronger assumptions on $\hat{\bm \Gamma}$ and $\hat{\bm \gamma}$ than prior sections. Specifically, Equation \ref{eq:summaryMR} assumes that $\hat{\bm \Gamma}$ and $\hat{\bm \gamma}$ are exactly normal and entries of the vector $(\hat{\bm \Gamma}, \hat{\bm \gamma})$ are independent of each other. This is stronger than Equation \ref{reduced form conv} where there can be dependence among $(\hat{\bm \Gamma}, \hat{\bm \gamma})$, and $(\hat{\bm \Gamma}, \hat{\bm \gamma})$ only have to be asymptotically normal.

\section{NONLINEAR MODELS}  \label{sec:nonlinearX}
Recent works have utilized non-linearities in the exposure model to identify and estimate $\beta$ in the presence of invalid instruments. The main idea is to leverage non-linear trends in the exposure model to create new instruments, which are then used to identify and estimate the causal effect of the exposure. Compared to linear methods, nonlinear treatment methods enable causal identification even when the plurality or majority assumptions are violated. But, as mentioned in Section \ref{sec:intro}, investigators should verify that the exposure model is indeed non-linear to ensure that these methods yield valid results. In this section, we review two recent works in this area.

First, \citet{guo2022two} considered the following modifications of Equation \ref{eq:DLinModel}:
\begin{align*}
D&=f(\mathbf{Z})+\delta, && \mathbb{E}(\delta \mid \mathbf{Z})=0, {\rm Var}[\mathcal{P}_{\mathbf{Z}}^\perp f(\mathbf{Z})] > 0,
 \\
& &&
\mathbb{E}[f(\mathbf{Z})] = 0, \mathbb{E}(\mathbf{Z}) = \mathbf{0}. \nonumber
\end{align*}
The term $\mathbf{Z}$ is a $p$ dimensional random variable and the term $\mathcal{P}_{\mathbf{Z}}^\perp f(\mathbf{Z}) = f(\mathbf{Z}) - \mathbf{Z}\tilde{\bm \gamma}$, $\tilde{\bm \gamma} = {\rm argmin}_{\bm \gamma} \mathbb{E}[f(\mathbf{Z}) - \mathbf{Z} {\bm \gamma}]^2$, is the residual from the best linear approximation of $f(\mathbf{Z})$. The positive variance assumption ${\rm Var}[\mathcal{P}_{\mathbf{Z}}^\perp f(\mathbf{Z})] > 0$
ensures that $f(\mathbf{Z})$ is a non-linear function of $\mathbf{Z}$. More generally, the positive variance assumption states that the exposure $D$ can be explained through linear and non-linear functions of $\mathbf{Z}$. In contrast, the outcome's relationship with the instruments in Equation \ref{eq:YLinModel} is linear. Critically, this discrepancy allows opportunities to create a ``non-linear'' instrument and identify $\beta$ with invalid instruments. To put it differently, the nonlinearity assumption on $f$ guarantees that the association between the treatment and the instrument, which is also characterized by the function $f(\cdot)$, is more complicated than the functional form of the violation, which is linear. 
We can formalize this observation by noticing that at the true value of $\beta$, the following equation holds:
\[
\mathbb{E}\{ [\mathcal{P}_{\mathbf{Z}}^\perp {f}(\mathbf{Z})]\cdot(\mathbf{Y} - \mathbf{D} \beta) \} = \mathbb{E}\{ [\mathcal{P}_{\mathbf{Z}}^\perp {f}(\mathbf{Z})]\cdot(\mathbf{Z}^{\intercal} \bm \pi) \}=0.
\]
The first equality follows from $\mathbb{E}[\epsilon \mid \mathbf{Z}] = 0$ in Equation \ref{eq:YLinModel}, and the second equality follows from the property of the orthogonal projection $\mathcal{P}_{\mathbf{Z}}^\perp$. \citet{guo2022two} also discusses a generalization of the above observation to the case where the instruments' direct effect on the outcome is non-linear. 

For estimation, \citet{guo2022two} proposed the following two-step method with sample splitting. Suppose we randomly split the data into two parts where the data in the first part is indexed as $i=1,\ldots,n_1$ and the data in the second part is indexed as $i=n_1+1,\ldots,n$. In the first step, $f(\mathbf{Z})$ is estimated using a nonparametric estimator or a machine learning algorithm (e.g., random forests) based on the data from the second part. In the second step, the fitted function $f$ is evaluated in the data from the first part and \citet{guo2022two} showed that this fit can be represented as
\[
\begin{bmatrix}
\hat{f}(\mathbf{Z}_{1}) \\
\vdots \\
\hat{f}(\mathbf{Z}_{n_1})
\end{bmatrix} =
\mathbf{Q} \tilde{\mathbf{D}}, \quad{} \tilde{\mathbf{D}} = \begin{pmatrix} D_{1} \\
\vdots \\
D_{n_1}
\end{pmatrix} \in \mathbb{R}^{n_1 \times 1}, \quad{} \mathbf{Q} \in \mathbb{R}^{n_1 \times n_1}.
\]
The matrix $\mathbf{Q}$ can be thought of as a matrix representation of the nonparametric estimator used in the second part of the data. For example, if $f(\mathbf{Z})$ is estimated via split random forests, each row of the matrix $\mathbf{Q}$ represents a $n_1$-dimensional aggregation weight \citep{lin2006random,meinshausen2006quantile,wager2018estimation}
. Let
\[
\tilde{\mathbf{Y}} = \mathbf{Q}\begin{pmatrix}
Y_{1} \\
\vdots \\
Y_{n_1}
\end{pmatrix} \in \mathbb{R}^{n_1 \times 1}, \quad{}  \tilde{\mathbf{Z}} = \mathbf{Q}\begin{pmatrix} \mathbf{Z}_{1} \\
\vdots \\
\mathbf{Z}_{n_1}
\end{pmatrix} \in \mathbb{R}^{n_1 \times p}, \quad{}  \mathbf{M} = \mathbf{Q}^{\intercal} \mathbf{P}_{\tilde{\mathbf{Z}}}^\perp\mathbf{Q} \in \mathbb{R}^{n_1 \times n_1}.
\]
Then, we introduce the following bias-corrected estimator of $\beta$,  
\begin{equation}
\hat{\beta}_{\rm tsci} =\tilde{\beta} - \frac{\sum_{i=1}^{n_1} M_{i,i} \hat{\delta}_i \hat{\epsilon}_i}{\tilde{\mathbf{D}}^\intercal \mathbf{M} \tilde{\mathbf{D}}}, \quad{} 
\tilde{\beta} = \frac{{\mathbf{Y}}^\intercal \mathbf{M} \mathbf{D}}{{\mathbf{D}}^\intercal \mathbf{M} {\mathbf{D}}},
\label{eq: ML corrected hetero}
\end{equation}
where $\hat{\delta}_{i} = D_i - \hat{f}(\mathbf{Z}_i)$ and $\hat{\epsilon}_i$ is the $i$th element of the vector $\mathbf{P}_{\mathbf{{Z}}}^\perp ({\mathbf{Y}} - {\mathbf{D}} \tilde{\beta})$. Because identification of $\beta$ relies on the nonlinear curvature $\mathcal{P}_{\mathbf{Z}}^\perp {f}(\mathbf{Z})$ and the estimation of $\beta$ uses a two-step procedure,  the estimator $\hat{\beta}_{\rm tsci}$ is referred to as ``Two-Stage Curvature Identification'' (TSCI) estimator.

Second, \cite{sun2023semiparametric} proposed to use 
higher-order interactions of $p$ instruments in the exposure model to
identify $\beta$. Similar to \citet{guo2022two}, \citet{sun2023semiparametric} generated ``new'' instruments from the $p$ instruments and the new instruments have non-linear effects on the exposure. However, \citet{guo2022two} used non-parametric estimators to create these new instruments whereas \citet{sun2023semiparametric} used higher-order interactions to create them. A bit more formally, suppose all $p$ instruments are mutually independent and there is at least $v$ valid instruments (i.e., $|\mathcal{V}| \geq v$ and $v \geq 1$); see \citet{sun2023semiparametric} when the instruments are dependent. Then, using the G estimation framework \citep{robins1992estimating,robins1994correcting}, \citet{sun2023semiparametric} showed that there is a function $\mathbf{h}^{[v]}(\mathbf{Z}) \in \mathbb{R}^d$ with $d = \sum_{j=0}^{v-1} {p \choose j}$ such that $\beta$ is the unique solution to 
\begin{equation} \label{eq:sun_esteq}
\mathbb{E}[\mathbf{h}^{[v]}(\mathbf{Z})(Y-\beta D)]= \mathbf{0}, \quad{} |\mathcal{V}| \geq v, 
\end{equation}
if $\mathbf{h}^{[v]}(\mathbf{Z})$ is associated with the exposure $D$. The function $\mathbf{h}^{[v]}(\mathbf{Z})$ represents all higher-order interactions of $p$ instruments of order greater than or equal to $v$. Specifically, for each $j=0,\ldots,v-1$, we create all possible subsets of $p$ instruments of size $p-j$ and denote this set as $\mathcal{C}_j$: 
\begin{align*}
\mathcal{C}_{0} &= \{ (1,\ldots,p) \}, \\
\mathcal{C}_{1} &= \{ (1,\ldots,p-1), (1,\ldots,p-2,p),...,(2,\ldots,p)\}, \\
\mathcal{C}_{j} &= \{ C \subseteq \{1,\ldots,p\} \mid |C| = p-j\}, \quad{} j=0,\ldots,v-1.
\end{align*}
Then, for each element $C \in \mathcal{C}_j$, we create the interaction instrument $\prod_{k\in C} [Z_j - \mathbb{E}(Z_j)]$. For example, for $\mathcal{C}_0$ and $\mathcal{C}_1$, we have the following interaction instruments:
\begin{align*}
\mathcal{C}_{0}&\Rightarrow \prod_{k=1}^{p} [Z_k - \mathbb{E}(Z_k)], \\
\mathcal{C}_{1}&\Rightarrow \prod_{k\neq p}[Z_k - \mathbb{E}(Z_k)], \prod_{k\neq p-1}[Z_k - \mathbb{E}(Z_k)], \ldots, \prod_{k\neq 1} [Z_k - \mathbb{E}(Z_k)].
\end{align*}
Stacking all the interaction instruments generated by every $\mathcal{C}_j, j=0,\ldots,v-1$ into a vector defines the function $\mathbf{h}^{[v]}(\mathbf{Z})$. Or, in other words, $\mathbf{h}^{[v]}(\mathbf{Z})$ creates $d$ interaction instruments.

As an illustrative example, suppose we have $p=2$ instruments $\mathbf{Z} = (Z, Z')$, $Z$ and $Z'$ are mutually independent, and at least one of them is valid (i.e. $v=1$). Then, $d = \sum_{j=0}^{v-1} {p \choose j} = 1$, $\mathcal{C}_0 = \{\{1,2\}\}$, and $h^{[1]}(\mathbf{Z}) = [Z - \mathbb{E}(Z)][Z' - \mathbb{E}(Z')]$. As long as $[Z - \mathbb{E}(Z)][Z' - \mathbb{E}(Z')]$ is associated with $D$, $\beta$ is the unique solution of Equation \ref{eq:sun_esteq} because
\[
\mathbb{E}[(Z - \mathbb{E}[Z])(Z' - \mathbb{E}[Z']) (Y- \beta D)] = \mathbb{E}[(Z - \mathbb{E}[Z])(Z' - \mathbb{E}[Z']) Z_{\mathcal{V}^C} \pi_{\mathcal{V}^C}] = 0.
\]
From the law of total expectations, the last equality holds for any $\mathcal{V}$ so long as $|\mathcal{V}| \geq v=1$ and the two instruments are independent of each other. Also, the last equality continues to hold for any $\mathcal{V}$ with $|\mathcal{V}|\geq 1$ even if the term $Z_{\mathcal{V}^{C}} \pi_{\mathcal{V}^c}$ in Equation \ref{eq:YLinModel} is non-linear, for instance if $Z_{\mathcal{V}^{C}} \pi_{\mathcal{V}^c}$
is replaced by an unknown function $\pi(Z_{\mathcal{V}^c})$. Or more loosely stated, in addition to $Z$ and $Z'$, $h^{[1]}(\mathbf{Z})$ serve as the ``new'', interaction instrument and so long as one of these instruments are valid, we can still identify the parameters in Equation \ref{eq:YLinModel}. 

For estimation, we can replace the terms in Equation \ref{eq:sun_esteq} with their empirical counterparts. Or, we can run two-stage least squares with the interaction instruments $\mathbf{h}^{[v]}(\mathbf{Z})$ and the term $\mathbb{E}(Z_j)$ in $\mathbf{h}^{[v]}(\mathbf{Z})$ is replaced by the $j$th column mean of the instrument matrix $\mathbf{Z} \in \mathbf{R}^{n \times p}$. In the example above where we have two instruments and $v= 1$, the estimator simplifies to 
\[
 \hat{\beta}_{\rm g}^{[1]} = \frac{\sum_{i=1}^{n} (Z_i - \bar{Z})(Z_i' - \bar{Z}')Y_i}{\sum_{i=1}^{n} (Z_i - \bar{Z})(Z_i' - \bar{Z}')D_i}
\]
The terms $\bar{Z}$ and $\bar{Z}'$ represent the mean of $Z$ and $Z$', respectively. 
The statistical properties of $\hat{\beta}_{\rm g}^{[v]}$ can be established by using the M-estimation framework. We remarked that \citet{sun2023semiparametric} also proposed a  new multiply robust identification framework, and a semiparametric efficient estimator of $\beta$.

\section{HETEROSKEDASTIC MODELS} \label{sec:GENIUS}
Another approach to study the causal effect of the exposure with invalid instruments is by leveraging heteroskedasticity of the observed data \citep{lewbel2012using,tchetgen2017genius,sun2022selective,ye2021genius}. Specifically, consider the following variations of Equations \ref{eq:DLinModel} and \ref{eq:YLinModel}:
\begin{align}
	& \mathbb{E}(Y\mid D,  \mathbf{Z}, U)=\beta D + \mathbf{Z} \bm \pi +\xi_y(U), \label{eq: out model}\\
	& \mathbb{E}(D\mid  \mathbf{Z},U)= \mathbf{Z} \bm\gamma + \xi_d(U), \quad{} \mathbf{Z} \perp U \label{eq: exp model}
\end{align}
where $\xi_y, \xi_d$ are unspecified functions. The variable $U$ represents an unmeasured variable that affects both the outcome $Y$ and the exposure $D$. \cite{tchetgen2017genius} showed that under Equations \ref{eq: out model} and \ref{eq: exp model}, $\beta$ can be identified as the unique solution to the following estimating equation
\begin{equation} \label{eq:genius}
\mathbb{E}\{ [\mathbf{Z}- \mathbb{E}(\mathbf{Z})][D - \mathbb{E}(D | \mathbf{Z})] (Y- \beta D)\}=\mathbf{0}
\end{equation}
as long as $D$ is heteroscedastic, i.e., $ {\rm Var}(D|\mathbf{Z}) $ varies as a function of  $\mathbf{Z}$. Specifically, under Equations \ref{eq: out model} and \ref{eq: exp model}, the left-hand side of Equation \ref{eq:genius} simplifies to
\begin{align*}
&\mathbb{E}\{ [\mathbf{Z}- \mathbb{E}(\mathbf{Z})][D - \mathbb{E}(D | \mathbf{Z})] [\mathbb{E}(Y \mid D, \mathbf{Z}, U) - \beta D]\} \\
=& \mathbb{E}\{ [\mathbf{Z}- \mathbb{E}(\mathbf{Z})][D - \mathbb{E}(D | \mathbf{Z})][\mathbf{Z}\bm{\pi} + \xi_y(U)]\}\\
=& \mathbb{E}\{ [\mathbf{Z}- \mathbb{E}(\mathbf{Z})][D - \mathbb{E}(D | \mathbf{Z})] \xi_y(U)\}  \\
=& \mathbb{E}\{ [\mathbf{Z}- \mathbb{E}(\mathbf{Z})][\xi_d(U) - \mathbb{E}[\xi_d(U)]] \xi_y(U)\} \\
=&  \mathbb{E}[\mathbf{Z}- \mathbb{E}(\mathbf{Z})] \mathbb{E} \{ [\xi_d(U) - \mathbb{E}[\xi_d(U)]] \xi_y(U)\} \\
=& \mathbf{0}
\end{align*}
We remark that the above argument holds more broadly even if we replace $\mathbf{Z} \bm \pi$ in Equation \ref{eq: out model} with any function $\mathbf{Z}$ or if we replace $\mathbf{Z} \bm \gamma$ in Equation \ref{eq: exp model} with any function of $\mathbf{Z}$. \citet{tchetgen2017genius} provides more general conditions under which Equation \ref{eq:genius} holds and refers to this framework as ``G-Estimation under No Interaction with Unmeasured Selection'' (GENIUS).

Intuitively,  Equation \ref{eq:genius} constructs $p$ new interaction instruments of the form $[\mathbf{Z} - \mathbb{E}(\mathbf{Z})][D-\mathbb{E}(D \mid \mathbf{Z})]$, which is the product of the original candidate set of instruments $\mathbf{Z}$ and the residual $D - \mathbb{E}(D \mid \mathbf{Z})$. From the independence of $U$ and $\mathbf{Z}$, the residual $D - \mathbb{E}(D \mid \mathbf{Z})$ is a proxy for $\xi_d(U)$. Then, the interaction instruments are valid instruments in that since there are no interactions between $U$ and $\mathbf{Z}$ in Equation \ref{eq: out model}, the interaction instruments satisfy the no direct effect assumption in condition (A2). Also, the interaction instruments are correlated with the exposure where for any $j = 1,\ldots,p$, we have
\begin{align*}
&{\rm Cov}\{[Z_j - \mathbb{E}(Z_j)][D-\mathbb{E}(D \mid \mathbf{Z})],D\} \\
=& \mathbb{E}\{[Z_j - \mathbb{E}(Z_j)][D-\mathbb{E}(D \mid \mathbf{Z})]D\} \\
=& \mathbb{E}\{[Z_j - \mathbb{E}(Z_j)]{\rm Var}(D \mid \mathbf{Z})\} + \mathbb{E}\{[Z_j - \mathbb{E}(Z_j)][D - \mathbb{E}(D \mid \mathbf{Z})]\mathbb{E}(D \mid \mathbf{Z})\} \\
=& \mathbb{E}\{[Z_j - \mathbb{E}(Z_j)]{\rm Var}(D \mid \mathbf{Z})\} \neq 0.
\end{align*}
The first equality uses the definition of covariance and the second equality uses $D = D - \mathbb{E}[D \mid \mathbf{Z}] + \mathbb{E}[D \mid \mathbf{Z}]$ along with the definition of conditional variance. The third equality uses the law of total expectation that conditions on $\mathbf{Z}$ and the final inequality is due to heteroskedasticity of $D$. Combined, the interaction instruments $[\mathbf{Z} - \mathbb{E}(\mathbf{Z})][D-\mathbb{E}(D \mid \mathbf{Z})]$ are ``new'' instruments to identify $\beta$ in Equation \ref{eq: out model}. Note that this approach is similar to Section \ref{sec:nonlinearX} where higher-order interactions of $\mathbf{Z}$ are used to create new instruments and identify $\beta$. Also, because the above identification strategy relies on heteroskedasticity of the exposure $D$ to create the new interaction instruments, it's possible to identify $\beta$ in Equation \ref{eq: out model} even if all instruments have direct effects on the outcome, i.e., if $\pi_j \neq 0$ for all $j$.

For estimation, one simple approach is to solve the sample equivalent version of Equation \ref{eq:genius}:
\begin{equation*}
\hat{\beta}_{\rm genius} = \underset{ \beta }{\rm argmin } \sum_{i=1}^{n}
[(\mathbf{Z}_i - \bar{\mathbf{Z}})(D_i - \mathbf{Z}_i \hat{\bm \gamma})(Y_i - \beta D_i)]^\intercal [(\mathbf{Z}_i - \bar{\mathbf{Z}})(D_i - \mathbf{Z}_i \hat{\bm \gamma})(Y_i - \beta D_i)].
\end{equation*}
The estimator $\hat{\beta}_{\rm genius}$ replaces $\mathbb{E}[D \mid \mathbf{Z}]$ with an estimate from a linear regression model where we regress $D$ on $\mathbf{Z}$. Under some moment assumptions,
\citet{tchetgen2017genius} show that $\hat{\beta}_{\rm genius}$ is consistent and asymptotically normal for $\beta$. \citet{ye2021genius} presents an extension of this estimator that is robust to many weak  invalid instruments and \citet{sun2022selective} proposes multiply robust estimator of $\beta$ where they use machine learning estimators for estimating relevant nuisance functions (e.g., $E[D \mid \mathbf{Z}]$).

 Finally, we discuss another idea based based on heteroskedastcity by \citet{liu2023mendelian}. Following previous notations, the authors considered the following variation of Equation \ref{eq:YLinModel}:
 \begin{equation} \label{eq:misteri}
 Y = \beta_0 + \beta D + \mathbf{Z} \bm \pi + \alpha D \exp(\eta_0 + \mathbf{Z} \bm \eta) + \xi, \quad{} \xi \mid D, \mathbf{Z} \sim N\left[0, \exp(\eta_0 + \mathbf{Z} \bm \eta) \right], \bm \eta \neq \mathbf{0}.
 \end{equation}
\noindent Here, the parameter $\beta$ represents the average treatment effect on the treated and is the target parameter of interest. 
Equation \ref{eq:misteri} is the observable implication of the following identification assumptions: (a) no additive interaction between $D$ and $\mathbf{Z}$ (i.e., $\mathbb{E}[Y(d,\mathbf{z}) - Y(0,\mathbf{z}) \mid D=d, \mathbf{Z} = \mathbf{z}] = d\beta$ for all $\mathbf{z}$), (b) homogeneous confounding of $D$; and (c) the outcome $Y$ among $D=0$ following a normal distribution  with variance $\exp(\eta_0 + \mathbf{Z} \bm \eta)$ and $\bm \eta \neq \mathbf{0}$. 
Condition (a) is weaker than the constant effect assumption where $Y(d,\mathbf{z}) - Y(0,\mathbf{z})  = d\beta$ and is satisfied if $\mathbf{Z}$ does not modify the average treatment effect on the treated.
Condition (b) is defined on the odds ratio scale and encodes the assumption that confounding on $D$ does not depend on $\mathbf{Z}$; see \citet{liu2023mendelian} for additional discussions. Conditions (b) and (c) give rise to the term $\alpha D \exp(\eta_0 + \mathbf{Z} \bm\eta) + \xi$ in Equation \ref{eq:misteri}. The distributional assumption can be relaxed to a mixture of normal distributions or to a distribution of $Y$ given $D=0$ that is heteroskedsatic in $\mathbf{Z}$. Finally, estimation of $\beta$ is based on the likelihood principle where we maximize the log likelihood of $\mathbb{P}(Y_i \mid D_i, \mathbf{Z}_i)$ in Equation \ref{eq:misteri}; we denote this estimator $\hat{\beta}_{\rm misteri}$ where MiSTERI stands for ``Mixed-Scale Treatment Effect Robust Identification.'' Another estimation approach based on the method of moments is discussed in \citet{liu2023mendelian}.

\section{ILLUSTRATION WITH REAL DATA}
\subsection{Background and Setup}
 We demonstrate the methods introduced above by re-analyzing the effect of body mass index (BMI) on systolic blood pressure (SBP) from the UK Biobank; see Section 5 of \citet{sun2023semiparametric} for details. Briefly, the UK Biobank is a large-scale prospective cohort study that recruited roughly 500,000 participants between 2006 and 2010 in the United Kingdom \citep{sudlow2015uk}. In the dataset, BMI was measured in units of kilograms per meter squared and SBP was measured in units of millimeters of mercury. Following \citet{sun2023semiparametric}, we restrict our analysis to people of genetically verified white British descent \citep{Tyrrell2016} and who are not taking anti-hypertensive medication based on self reporting. The sample size for the final analysis is $n=292,757$. We use the top $p=10$ single nucleotide polymorphisms (SNPs) ranked by their $p$-values, each of which were derived from testing the effect of a SNP on BMI with simple linear regression. The 10 p-values reach genome-wide significance level of $5\times 10^{-8}$ \citep{Locke:2015aa} and have pairwise correlation coefficients that are less than 0.01.  The 10 SNPs are rs1558902,  rs6567160,   rs543874, rs13021737, rs10182181,  rs2207139, rs11030104, rs10938397, rs13107325, and rs3888190. 
 The overall F-statistic for the first-stage model (i.e., Equation \ref{eq:DLinModel} with 10 instruments) is 146.1, with a p-value that is less than 10$^{-8}$. To focus on problems caused by invalid instruments, we chose the top 10 SNPs here to minimize effects from weak instruments; see \cite{sun2023semiparametric} for further details on how these instruments were chosen.
 

 We compare the following methods for estimating $\beta$: $\hat{\beta}_{\rm med}$, $\hat{\beta}_{\rm sisvive}$, $\hat{\beta}_{\rm kclass}$ $\hat{\beta}_{\rm adlasso}$, $\hat{\beta}_{\rm tsht}$, $\hat{\beta}_{\rm cim}$, $\hat{\beta}_{\rm tsci}$, $\hat{\beta}_{\rm g}^{[v]}$, $\hat{\beta}_{\rm genius}$, and $\hat{\beta}_{\rm misteri}$.  For $\hat{\beta}_{\rm g}^{[v]}$, we set the minimum number of valid instruments to be $6$ and $8$ and they are denoted as $\hat{\beta}_{\rm g}^{[6]}$ and $\hat{\beta}_{\rm g}^{[8]}$, respectively. We also compute he union confidence interval ${\rm CI}_{\rm union}^{[v]}$ and the search and sampling confidence interval ${\rm CI}_{\rm ss}$. For ${\rm CI}_{\rm union}^{[v]}$, we set the minimum number of valid instruments to be $v=6$ and $v=8$ and use the conditional likelihood ratio test \citep{moreira_conditional_2003}. 
 Finally, we include two baseline analyses of the causal effect. The first baseline analysis is the ordinary least squares estimator of $\beta$ that fits a linear regression of SBP on BMI. The second baseline analysis is the TSLS estimator of $\beta$ from Section \ref{sec:point_est} that sets $\mathcal{V}$ to be all 10 SNPs; in other words, this estimator assumes that all 10 SNPs are valid. 
 
We use the following software to compute the estimates or confidence intervals of $\beta$. To compute $\hat{\beta}_{\rm sisvive}$, we use the R package \texttt{sisVIVE} \citep{sisVIVE}. To compute $\hat{\beta}_{\rm kclass}$, we use the R package \texttt{ivmodel} \citep{kang2021ivmodel}. To compute $\hat{\beta}_{\rm med}$ and $\hat{\beta}_{\rm adlasso}$, we use the code provided in \citet{windmeijer_lasso_2019}. To compute $\hat{\beta}_{\rm tsht}$ and the search and sampling confidence interval ${\rm CI}_{\rm ss}$, we use the R package \texttt{RobustIV} \citep{koo2023robustiv}. To compute the union confidence interval ${\rm CI}_{\rm union}^{[v]}$, we use the code provided in \citet{kang2022two}. To compute $\hat{\beta}_{\rm cim}$, we use the code provided in \citet{windmeijer_confidence_2021}.
To compute $\hat{\beta}_{\rm tsci}$, we use R package \texttt{TSCI} \citep{carl2023tsci}. To compute $\hat{\beta}_{\rm g}^{[v]}$, we use the code in the R package \texttt{MRSquare}.
Finally, to compute $\hat{\beta}_{\rm genius}$ and $\hat{\beta}_{\rm misteri}$, we use the code provided in \citet{tchetgen2017genius} and \citet{liu2023mendelian}, respectively.


\subsection{Results}
\begin{figure}[ht]

      \includegraphics[scale=0.4]{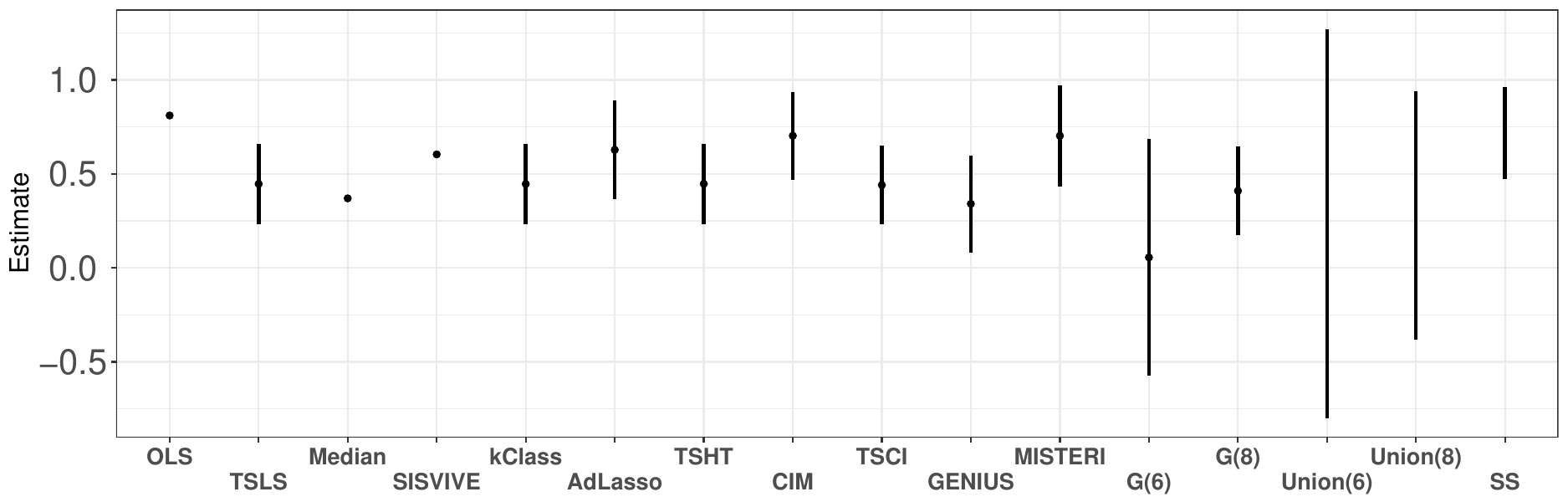}
      \caption{Point estimates (denoted as points) or 95\% confidence intervals (denoted as an interval) for the effect of body mass index (BMI) on systolic blood pressure (SBP) from the UK Biobank. Sample size is $n=292,757$ and there are $p=10$ genetic instruments. OLS refers to the ordinary least squares estimator of $\beta$ without using instruments. TSLS refers to $\hat{\beta}_{\rm tsls}$ that assumes all 10 instruments are valid. Median refers to $\hat{\beta}_{\rm med}$, SISVIVE refers to $\hat{\beta}_{\rm sisvive}$, k-class refers to $\hat{\beta}_{\rm kclass}$, AdLasso refers to $\hat{\beta}_{\rm adlasso}$, TSHT refers to $\hat{\beta}_{\rm tsht}$, CIM refers to $\hat{\beta}_{\rm cim}$, TSCI refers to $\hat{\beta}_{\rm tsci}$, GENIUS refers to $\hat{\beta}_{\rm genius}$, and MiSTERI refers to $\hat{\beta}_{\rm misteri}$. G(6) and G(8) refer to $\hat{\beta}_{\rm g}^{[6]}$ and $\hat{\beta}_{\rm g}^{[8]}$, respectively. SS refers to ${\rm CI}_{\rm ss}$. Also, ${\rm CI}_{\rm union}^{[6]}$ and ${\rm CI}_{\rm union}^{[8]}$ refer to union confidence intervals which assume at least 6 and 8 instruments are valid, respectively.}
      \label{fig:bmi-sbp}
  \end{figure}

The point estimates and 95\% confidence intervals for different methods are summarized in Figure \ref{fig:bmi-sbp}. Except for $\hat{\beta}_{\rm g}^{[6]}$, ${\rm CI}_{\rm union}^{[6]}$, and ${\rm CI}_{\rm union}^{[8]}$, all methods, including the TSLS estimator that assumes all instruments are valid and the OLS estimator that does not use any instruments, suggest that there is a positive effect of increasing BMI on SBP. The largest and smallest values of the causal effect are from the union confidence interval that assumes at least 6 instruments are valid; its upper confidence limit of the causal effect is $1.2710$ and its lower confidence limit is $-0.799$. The G-estimator that assumes at least six instruments are valid (i.e. $\hat{\beta}_{\rm g}^{[6]}$) also gives wide confidence intervals, ranging from $-0.5740$ to $0.6860$. But after we increase the number of valid IVs to 8, the confidence interval is $(0.174,  0.646)$ and no longer covers $0$. In general, the G-estimator and the union confidence interval allow users to conduct sensitivity analysis by varying the number of valid IVs, and can be used to reflect the uncertainty about the validity of instruments.

Among methods that generate confidence intervals, their 95\% confidence intervals overlap with each other. In other words, after accounting for sampling uncertainty, these methods, despite making different assumptions about instrument invalidity, do not differ from each other with respect to their conclusions about the causal effect. Excluding the OLS estimator, the narrowest 95\% confidence interval is generated from $\hat{\beta}_{\rm TSCI}$ and the widest confidence interval is generated from the union confidence interval that assumes at least six instruments are valid.

Despite almost all methods suggesting that the effect of BMI on SBP is positive, we do see that the methods roughly cluster into two types. The first cluster of methods (i.e., SISVIVE, the adaptive Lasso, the confidence interval method, MiSTERI, and the search and sampling confidence interval) roughly estimates the causal effect to be around $0.68$ and they are close to the OLS estimator that does not use any instruments. The second cluster of methods (i.e., the median estimator, the k-class estimator, TSHT, TSCI, the G estimators, GENIUS, and the union confidence intervals) roughly estimates the causal effect to be around $0.40$ and they are close to the TSLS estimator that assumes all 10 instruments are valid.

Among methods that select valid instruments, CIM selected rs543874 and rs10182181 as invalid instruments. TSHT selected rs10182181 as an invalid instrument. The adaptive Lasso selected rs543874, rs10182181, and rs13107325 as invalid instruments. 
Across all methods that can select valid instruments,  instrument rs10182181 was selected as an invalid instrument.

\section{DISCUSSION}

This paper provides a review of identification and inference of the causal effect of the exposure on the outcome when there are invalid instruments. We start with the linear model framework where the parameter $\bm \pi$ in Equation \ref{eq:YLinModel} encodes the violations of the IV assumptions. Broadly speaking, works in this framework require either that the majority of instruments or a plurality of instruments are valid to obtain identification and inference of $\beta$. Subsequent works have leveraged non-linearities or heteroskedastcities in the models to identify and infer $\beta$. 
In our data analysis, we find that all methods yield similar conclusions about the effect of body mass index on systolic blood pressure. 

Despite significant progress in identifying and inferring the causal effect of the exposure in the presence of invalid instruments, several challenges remain and we highlight a couple of them. First, as illustrated throughout the paper, there are different ways to define a valid (or an invalid) instrument based on how the instrument deviates from the assumptions (A2) and (A3). Roughly speaking, Section \ref{sec:linmodel} defines an invalid instrument through a ``linear deviation'' from the IV assumptions (A2) and (A3). In contrast, Sections \ref{sec:nonlinearX} and \ref{sec:GENIUS} defines an invalid instrument through both linear and ``non-linear'' deviations from the IV assumption. Because the latter sections allow for broader types of invalid instruments, they typically require additional conditions on the data-generating model for identification and inference, such as a non-linear exposure model or a heteroskedastic exposure or outcome model. Second, while methods for uniform inference exist, there is still room for improvement, especially compared to the oracle TSLS estimator that knows which instruments are valid a priori. Third, only a handful of works have explored how to conduct valid inference when instruments are both invalid and are weakly associated with the exposure; these instruments are common in Mendelian randomization where the genetic variants only explain a fraction of the variance in the exposure and most of them are suspected to be pleiotropic. \citet{guo_confidence_2018} proposes a thresholding procedure in TSHT to select instruments that are strongly correlated with the exposure. \cite{lin2024instrumental} considers many weak instruments under the plurality rule where instead of only using instruments that are strongly correlated with the exposure, they use both strong and weak instruments. \cite{zhang_fighting_2022} considered another approach to select strong instruments that also prevents the mis-selection of valid instruments. \citet{ye2021genius} proposes an improved version of the GENIUS estimator that allows for many weak invalid instruments discussed in \citet{newey2009generalized}. The method in \citet{kang2022two} allows for uniformly valid inference in the presence of weak instruments defined by \citet{staiger_instrumental_1997} and invalid instruments defined in Definition \ref{def:validIV}. Fourth, there is still an open question about how to select the optimal set of instruments for  bias reduction and/or efficiency improvement when invalid instruments are present. This problem is especially challenging when faced with many weak instruments \citep{liu2023mendelian,ye2021genius,lin2024instrumental}.




\section*{DISCLOSURE STATEMENT}
The authors are not aware of any affiliations, memberships, funding, or financial holdings that
might be perceived as affecting the objectivity of this review. 

\section*{ACKNOWLEDGMENTS}
Dr. Zijian Guo’s research was supported in part by NIH grants R01-GM140463 and R01-LM013614. Dr. Zhonghua Liu's research was supported in part by NIH grant 1R01AG086379-01.
Dr. Dylan Small's research was supported in part by NIH grant 5R01AG065276-02. 

\bibliographystyle{ar-style1}
\bibliography{mainbib}

\end{document}